\pdfoutput=1

\documentclass[11pt,twoside,a4paper,cmspaper,final,collab]{cms-tdr}
\def\svnVersion{266932}\def\cmsCernNo{2013-037}\def\cmsCernDate{\today}\def\cmsMessage{Published in the Journal of High Energy Physics as \href{http://dx.doi.org/10.1007/JHEP09(2012)094}{\doi{10.1007/JHEP09(2012)094.}}}

\begin{document}\cmsNoteHeader{EXO-11-059}

\hyphenation{had-ron-i-za-tion}
\hyphenation{cal-or-i-me-ter}
\hyphenation{de-vices}
\RCS$Revision: 132457 $
\RCS$HeadURL: svn+ssh://svn.cern.ch/reps/tdr2/papers/EXO-11-059/trunk/EXO-11-059.tex $
\RCS$Id: EXO-11-059.tex 132457 2012-06-25 12:20:37Z alverson $
\newcommand{\todo}[1]{{\color{magenta}{FIXME: #1}}}
\renewcommand{\fixme}[1]{\todo{#1}}
\newcommand{\sectionauthor}[1]{\texorpdfstring{{\color{blue} #1}}{#1}}
\newcommand{\disclaimer}[1]{\emph{Outlook: #1}}
\newcommand{\update}[1]{{\color{blue}{#1}}}

\newcommand{\T}{\ensuremath{\,\text{T}}\xspace}                
\newcommand{\meter}{\ensuremath{\,\text{m}}\xspace}            
\newcommand{\um}{\mu\ensuremath{\,\text{m}}\xspace}            
\newcommand{\ns}{\ensuremath{\,\text{ns}}\xspace}              
\newcommand{\us}{\mu\ensuremath{\,\text{s}}\xspace}            
\newcommand{\mrad}{\ensuremath{\,\text{mrad}}\xspace}          
\newcommand{\urad}{\mu\ensuremath{\,\text{rad}}\xspace}        
\newcommand{\kHz}{\ensuremath{\,\text{kHz}}\xspace}            
\newcommand{\MHz}{\ensuremath{\,\text{MHz}}\xspace}            
\newcommand{\GHz}{\ensuremath{\,\text{GHz}}\xspace}            
\newcommand{\rad}{\ensuremath{\,\text{rad}}\xspace}            
\newcommand{\pb}{\ensuremath{\,\text{pb}}\xspace}

\def\antibar#1{\ensuremath{#1\bar{#1}}}
\newcommand{\ppbar}{\antibar{\mathrm{p}}\xspace}
\newcommand{\pp}{\ensuremath{\mathrm{pp}}\xspace}
\newcommand{\e}{\ensuremath{\mathrm{e}}\xspace}
\newcommand{\epm}{\ensuremath{\mathrm{e^{\pm}}}\xspace}
\newcommand{\W}{\ensuremath{\mathrm{W}}\xspace}
\newcommand{\DM}{\ensuremath{\chi}\xspace}

\newcommand{\pti}[1]{\ensuremath{p_{\text{T},#1}}\xspace
}
\newcommand{\ptgen}{\ensuremath{\pt^{\text{gen}}}\xspace}
\newcommand{\pttrue}{\ensuremath{\pt^{\text{true}}}\xspace}
\newcommand{\ptmin}{\ensuremath{\pt^{\text{min}}}\xspace}
\newcommand{\ptmax}{\ensuremath{\pt^{\text{max}}}\xspace}
\newcommand{\ptreco}{\ensuremath{\pt(\text{recoJet})}\xspace}
\newcommand{\ptave}{\ensuremath{\pt^{\text{ave}}}\xspace}
\newcommand{\ptavemin}{\ensuremath{\pt^{\text{ave,min}}}\xspace}
\newcommand{\ptavemax}{\ensuremath{\pt^{\text{ave,max}}}\xspace}
\newcommand{\ptref}{\ensuremath{\pt^{\text{ref}}}\xspace}
\newcommand{\ppl}{\ensuremath{p_{||}}\xspace}
\newcommand{\ppli}[1]{\ensuremath{p_{||,#1}}\xspace}
\newcommand{\dif}[1]{\ensuremath{\text{d}#1}\xspace}
\newcommand{\mean}[1]{\ensuremath{\langle#1\rangle}}
\newcommand{\tev}{\TeV}
\newcommand{\mht}{\ensuremath{\slash\mkern-12mu{H}_{\text{T}}}\xspace}

\newcommand{\pthat}{\ensuremath{\hat{\text{p}}_\mathrm{T}}\xspace}
\newcommand{\ptrecjet}{\ensuremath{p_{\mathrm{T}}^{\mathrm{recoJet}}}\xspace}
\newcommand{\ptparjet}{\ensuremath{p_{\mathrm{T}}^{\mathrm{particleJet}}}\xspace}
\newcommand{\ptgenjet}{\ensuremath{p_{\mathrm{T}}^{\mathrm{GenJet}}}\xspace}
\newcommand{\ptg}{\ensuremath{p_{\mathrm{T}}^{\gamma}}\xspace}
\newcommand{\ptj}{\ensuremath{p_{\mathrm{T}}^{\mathrm{jet}}}\xspace}
\newcommand{\ptja}{\ensuremath{p_{\mathrm{T}}^{\mathrm{jet1}}}\xspace}
\newcommand{\ptjb}{\ensuremath{p_{\mathrm{T}}^{\mathrm{jet2}}}\xspace}
\newcommand{\ptjc}{\ensuremath{p_{\mathrm{T}}^{\mathrm{jet3}}}\xspace}
\newcommand{\rj}{\ensuremath{R_{\mathrm{jet}}}\xspace}
\newcommand{\rg}{\ensuremath{R_{\mathrm{G}}}\xspace}
\newcommand{\ptiv}[1]{\ensuremath{\mathbf{p}_{\mathrm{T},#1}}\xspace}
\newcommand{\ptivsq}[1]{\ensuremath{|{\mathbf{p}_{\mathrm{T},#1}|}^{2}}\xspace}
\newcommand{\ptivsqm}[1]{\ensuremath{|{\mathbf{p}_{\mathrm{T},#1}^{\mathrm{meas}}|}^{2}}\xspace}
\newcommand{\ptivsqt}[1]{\ensuremath{|{\mathbf{p}_{\mathrm{T},#1}^{\mathrm{true}}|}^{2}}\xspace}
\newcommand{\ptivm}[1]{\ensuremath{\mathbf{p}_{\mathrm{T},#1}^{\mathrm{meas}}}\xspace}
\newcommand{\ptivt}[1]{\ensuremath{\mathbf{p}_{\mathrm{T},#1}^{\mathrm{true}}}\xspace}
\newcommand{\ptjeta}{\ensuremath{\mathbf{p}_{\mathrm{T}}^{\mathrm{jet1}}}\xspace}
\newcommand{\ptjetb}{\ensuremath{\mathbf{p}_{\mathrm{T}}^{\mathrm{jet2}}}\xspace}
\newcommand{\ptjetc}{\ensuremath{\mathbf{p}_{\mathrm{T}}^{\mathrm{jet3}}}\xspace}
\newcommand{\ptjetd}{\ensuremath{\mathbf{p}_{\mathrm{T}}^{\mathrm{jet4}}}\xspace}
\newcommand{\metSM}{\ensuremath{E_{\mathrm{T}}^{\mathrm{Estimated}}}\xspace}
\newcommand{\metC}{\ensuremath{E_{\mathrm{T}}^{\mathrm{C}}}\xspace}
\newcommand{\metbf}{\ensuremath{{\mathbf{E}}_{\mathrm{T}}^{\mathrm{miss}}}\xspace}
\newcommand{\met}{\MET}

\newcommand\defRTDR{\ensuremath{ R2 = \sqrt{[\pi-\Delta\phi(J_{1},\mht) ]^2 +[\Delta\phi(J_{2},\mht) ]^2}    }}
\newcommand\wpj{\ensuremath{\W\textrm{+jets}}\xspace}
\newcommand\zpj{\ensuremath{\Z\textrm{+jets}}\xspace}
\renewcommand\ttbar{\ensuremath{{\rm t\bar{t}}}\xspace}
\newcommand\ttbarMuNu{\ensuremath{\ttbar (\rightarrow \mu \nu) {\rm +jets}}\xspace}
\newcommand\wtauhad{\ensuremath{\W \rightarrow \tau_{\rm h} \nu}\xspace}
\newcommand\ttbarTAUHNu{\ensuremath{\ttbar \rightarrow \tau_{\rm h} \nu + {\rm jets}}\xspace}
\newcommand\ttbarDITAUHNu{\ensuremath{\ttbar \rightarrow \tau_{\rm h} \nu + \tau_{\rm h} \nu + {\rm jets}}\xspace}
\newcommand\ttbarMUTAUHNu{\ensuremath{\ttbar (\rightarrow \mu \nu + \tau_{\mu} \nu) {\rm +jets}}\xspace}
\newcommand\defMHT{\ensuremath{ {\rm MHT} = \mht = | - \sum_{i} \vec{p_{T} (jet_{i}) }|}}
\newcommand{\ztautau}{\ensuremath{\Z \rightarrow \tau \tau}\xspace}
\newcommand{\zmumu}{\ensuremath{\Z \rightarrow \mu \mu}\xspace}
\newcommand{\zee}{\ensuremath{\Z \rightarrow \e \e}\xspace}
\newcommand{\znunu}{\ensuremath{\Z \rightarrow \nu \bar{\nu}}\xspace}
\newcommand{\zll}{\ensuremath{\Z \rightarrow \ell^{+}\ell^{-}}\xspace}
\newcommand{\wlnu}{\ensuremath{\W \rightarrow \ell\nu}\xspace}
\newcommand{\wenu}{\ensuremath{\W \rightarrow \e\nu}\xspace} 
\newcommand{\wmunu}{\ensuremath{\W \rightarrow \mu \nu}\xspace}
\newcommand{\zellell}{\ensuremath{\Z \rightarrow \ell^{+} \ell^{-}}\xspace}
\newcommand{\znunubr}{\ensuremath{\Z ( \rightarrow \nu \bar{\nu} )}\xspace}
\newcommand{\zellellbr}{\ensuremath{\Z ( \rightarrow \ell^{+} \ell^{-} )}\xspace}
\newcommand{\wmunubr}{\ensuremath{\W (\rightarrow \mu \nu )}\xspace}
\newcommand{\wenubr}{\ensuremath{\W (\rightarrow \e\nu )}\xspace}
\newcommand{\wlnubr}{\ensuremath{\W ( \rightarrow \ell \nu )}\xspace}
\newcommand{\wellnubr}{\ensuremath{\W ( \rightarrow \ell \nu )}\xspace}
\newcommand{\wtaunubr} {\ensuremath{\W ( \rightarrow \tau \nu )}\xspace}
\newcommand{\ztautaubr}{\ensuremath{\Z (\rightarrow \tau \tau)}\xspace}
\newcommand{\zmumubr}  {\ensuremath{\Z (\rightarrow \mu \mu)}\xspace}
\newcommand{\zeebr}    {\ensuremath{\Z (\rightarrow \e \e)}\xspace}

\newcommand{\susy}{{\sc susy}\xspace}
\newcommand{\squark}{\sQua}
\newcommand{\gluino}{\sGlu}
\newcommand{\supq}{\sUp}
\newcommand{\sdown}{\sDw}
\newcommand{\sstrange}{\ensuremath{\tilde{s}}\xspace}
\newcommand{\scharm}{\ensuremath{\tilde{c}}\xspace}
\newcommand{\sbottom}{\sBot}
\newcommand{\MEt}{\not\!\! E_{\mathrm{T}}}

\newcommand{\mhtv}[1]{\ensuremath{\slash\mkern-15mu{\vec{H}}_{\text{T}}^{#1}}\xspace}
\newcommand{\METv}{\ensuremath{\slash\mkern-12mu{\vec{E}}_{\text{T}}}\xspace}
\newcommand{\MHT}{\mht}
\newcommand{\MHTv}{\mhtv{}}
\newcommand{\MPT}{\mathrm{MPT}}
\newcommand{\seed}{\mathrm{seed}}
\newcommand{\reco}{\mathrm{reco}}
\newcommand{\gen}{\mathrm{gen}}
\newcommand{\soft}{\mathrm{soft}}
\newcommand{\true}{\mathrm{true}}
\newcommand{\smeared}{\mathrm{smeared}}
\newcommand{\ptcl}{\mathrm{particle}}
\newcommand{\recoJet}{\mathrm{recoJet}}
\newcommand{\genJet}{\mathrm{genJet}}
\newcommand{\ptSoft}{\ensuremath{p_{\mathrm{T,\soft}}}}
\newcommand{\ptSoftV}{\ensuremath{\vec{p}_{\mathrm{T,\soft}}}}
\newcommand{\recoSoft}{\ensuremath{\vec{p}_{\mathrm{T,\soft}}^{\,\reco}}}
\newcommand{\genSoft}{\ensuremath{\vec{p}_{\mathrm{T,\soft}}^{\,\ptcl}}}
\newcommand{\trueSoft}{\ensuremath{\vec{p}_{\mathrm{T,\soft}}^{\,\true}}}
\newcommand{\seedSoft}{\ensuremath{\vec{p}_{\mathrm{T,\soft}}^{\,\seed}}}
\newcommand{\ptV}{\ensuremath{\vec{p}_{\mathrm{T}}}}
\newcommand{\ppp}[3]{\ensuremath{p_{#1,#2}^{#3}}}
\newcommand{\ppt}[2]{\ppp{\mathrm{T}}{#1}{#2}}
\newcommand{\pptV}[2]{\ensuremath{\vec{p}_{\mathrm{T},#1}^{\,#2}}}
\newcommand{\ptInv}{\ptV^{\,\mathrm{invisible}}}
\newcommand{\ptAve}[1]{\langle\pt^{#1}\rangle}
\newcommand{\particleJet}{\mathrm{particleJet}}
\newcommand{\seedJet}{\mathrm{seedJet}}
\newcommand{\seedJets}{\mathrm{seedJets}}
\newcommand{\seedMHT}{\MHT^{\seed}}
\newcommand{\seedMHTv}{\mhtv{\seed}}
\newcommand{\seedHT}{\mathrm{seedHT}}
\newcommand{\partonHT}{\mathrm{partonHT}}
\newcommand{\Dphi}{\Delta\phi}
\newcommand{\DphiMPTMHT}{\Delta\phi(\MPT,\MHT)}
\newcommand{\DphiMHTjet}[1]{\Delta\phi(\MHT,\mathrm{jet\,1\mbox{-}#1})}
\newcommand{\minDphi}{\mathrm{min}\,\Delta\phi(\MHT,\mathrm{jet\,1\mbox{-}3})}
\newcommand{\ssoftT}{\sigma_{\mathrm{T}}^{\soft}}
\newcommand{\ssoftPhi}{\sigma_{\mathrm{\phi}}^{\soft}}
\newcommand{\E}[1]{\times10^{#1}}
\newcommand{\maxL}{\mathrm{max\mbox{-}}\rsL}
\newcommand{\rsN}{(\mathrm{R}+\mathrm{S})}
\newcommand{\effMu}{\varepsilon_{\mu}}
\newcommand{\fakeMu}{\ensuremath{\slash\mkern-12mu{\mu}}\xspace}
\newcommand{\effSplit}{\varepsilon_{\mathrm{split}}}
\newcommand{\ptRel}{\ensuremath{\pt^{\mathrm{rel}}}}
\newcommand{\ptMu}{\ensuremath{\ptV^{\,\mu}}}
\newcommand{\ptJet}{\ensuremath{\ptV^{\,\mathrm{jet}}}}
\newcommand{\footnoteremember}[2]{\footnote{#2}\newcounter{#1}\setcounter{#1}{\value{footnote}}}
\newcommand{\footnoterecall}[1]{\footnotemark[\value{#1}]} 

\newcommand{\minus}{\mbox{-}}
\newcommand{\plus}{\mbox{+}}
\newcommand{\genParticle}{\mathrm{genParticle}}
\newcommand{\minSeedPT}{p_{\mathrm{T,min}}^{\mathrm{seed}}\xspace}
\newcommand{\smearedMHT}{\MHT^{\smeared}}
\newcommand{\smearedMHTv}{\mhtv{\smeared}}
\newcommand{\jet}{\mathrm{jet}}
\newcommand{\rsL}{L}
\newcommand{\rsLjets}{\rsL_{\mathrm{jets}}}
\newcommand{\rsLsoft}{\rsL_{\mathrm{soft}}}
\newcommand{\rsLcooled}{L_{\mathrm{cooled}}}
\newcommand{\lnLjs}[2]{\minus\ln\rsL_{#1\mathrm{j}+#2\mathrm{s}}^{\max}}
\newcommand{\minLnL}{\min(-\ln\rsL)}
\newcommand{\fcool}{f_{\mathrm{cool}}}
\newcommand{\calo}{\mathrm{calo}}
\newcommand{\caloSoft}{\ensuremath{\vec{p}_{\mathrm{T,\soft}}^{\,\calo}}}
\newcommand{\bisector}{\hat{\eta}}
\newcommand{\along}{\hat{\xi}}
\newcommand{\rs}{\mathrm{R}\mbox{+}\mathrm{S}}
\newcommand{\softRes}{R_{\soft}}
\newcommand{\lfrac}[2]{l_{#1}^{#2}}
\newcommand{\lfracApprox}[2]{l_{#1}^{\prime#2}}
\newcommand{\mufrac}[2]{\mu_{#1}^{#2}}
\newcommand{\mufracApprox}[2]{\mu_{#1}^{\prime#2}}
\newcommand{\br}{\mathrm{BR}}
\newcommand{\brMu}{\br_{\mu}}
\newcommand{\res}{r}
\newcommand{\resLep}{\res_{\mathrm{lep}}}
\newcommand{\resHad}{\res_{\mathrm{had}}}
\newcommand{\resBad}{\res_{\mathrm{bad}}}
\newcommand{\resGood}{\res_{\mathrm{good}}}
\newcommand{\resLepApprox}{\resLep^{\prime}}
\newcommand{\resHadApprox}{\resHad^{\prime}}
\newcommand{\distLep}{h_{\mathrm{lep}}^{\prime}}
\newcommand{\distHad}{h_{\mathrm{had}}^{\prime}}
\newcommand{\gausGood}{\mathcal{G}_{\mathrm{good}}}
\newcommand{\fracBad}{w_{\mathrm{bad}}}
\newcommand{\fracGood}{w_{\mathrm{good}}}
\newcommand{\dope}[2]{d_{#1}^{#2}}
\newcommand{\dopeApprox}[2]{d_{#1}^{\prime#2}}
\newcommand{\dopeMu}{d_{n,\mu\mbox{-}\mathrm{tag}}}
\newcommand{\dijet}{\mathrm{dijet}}
\newcommand{\ndof}{\mathrm{ndof}}
\newcommand{\METvJ}{\ensuremath{\slash\mkern-12mu{\,\vec{J}}_{\text{T}}}\xspace}
\newcommand{\badFrac}{\ensuremath{f_{\mathrm{ECAL}}^{\mathrm{masked}}}\xspace}

\newcommand{\DeltaPhi}{\ensuremath{\Delta\phi_{\text{min}}}}

\providecommand{\cPq}{\ensuremath{\cmsSymbolFace{q}}} 
\providecommand{\cPg}{\ensuremath{\cmsSymbolFace{g}}} 
\providecommand{\cPG}{\ensuremath{\cmsSymbolFace{G}}} 
\providecommand{\cPU}{\ensuremath{\cmsSymbolFace{U}}} 
\providecommand{\cPZ}{\ensuremath{\cmsSymbolFace{Z}}} 
\providecommand{\cPW}{\ensuremath{\cmsSymbolFace{W}}} 
\providecommand{\cPge}{\ensuremath{\cmsSymbolFace{ge}}} 
\providecommand{\cPgn}{\ensuremath{\cmsSymbolFace{gn}}} 
\providecommand{\cPgm}{\ensuremath{\cmsSymbolFace{gm}}} 
\newcommand{\LambU}{\ensuremath{{\Lambda_\cmsSymbolFace{U}}}\xspace}
\newcommand{\dU}{\ensuremath{{d_\cmsSymbolFace{U}}}\xspace}
\newcommand{\Wmunu}{\ensuremath{\W (\mu \nu )}\xspace}
\newcommand{\Zmumu}{\ensuremath{\Z (\mu \mu)}\xspace}
\newcommand{\Zellell}{\ensuremath{\Z (\ell \ell)}\xspace}
\newcommand{\Znunu}{\ensuremath{\Z (\nu \bar{\nu} )}\xspace}
\newcommand{\ZnunuJets}{{\ensuremath{\Z (\nu \bar{\nu})}}+jets\xspace}
\newcommand{\ZellellJets}{{\ensuremath{\Z (\ell \ell    )}}+jets\xspace}
\newcommand{\ZJets}{{\ensuremath{\Z}}+jets\xspace}
\newcommand{\WJets}{{\ensuremath{\W}}+jets\xspace}
\newcommand{\WmunuJet}{{\ensuremath{\W (\mu \nu)}}+jets\xspace}

\graphicspath{{figures/}}

\cmsNoteHeader{EXO-11-059}
\title{\texorpdfstring{Search for dark matter and large extra dimensions in monojet events in $\Pp\Pp$ collisions at $\sqrt{s} = 7$\TeV}{Search for dark matter and large extra dimensions in monojet events in pp collisions at sqrt(s)= 7 TeV}}

\date{\today}
\abstract{
 A search has been made for events containing an energetic jet and an imbalance in transverse momentum using a data sample of pp collisions at a center-of-mass energy of 7\TeV.
This signature is common to both dark matter and extra dimensions models. The data were collected by the CMS detector at the LHC and correspond to an
 integrated luminosity of 5.0\fbinv. The number of observed events is consistent with the standard model expectation. Constraints on the
dark matter-nucleon scattering cross sections are determined for both spin-independent and spin-dependent interaction models.
For the spin-independent model, these are the most constraining limits for a
dark matter particle with mass below 3.5\GeVcc, a region
unexplored by direct detection experiments.
For the spin-dependent model, these are the most stringent constraints over the 0.1--200\GeVcc mass range.
The constraints on the Arkani-Hamed, Dimopoulos, and Dvali model parameter $M_\mathrm{D}$ determined as a function of the number of extra dimensions are
also an improvement over the previous results.
}

\hypersetup{%
pdfauthor={CMS Collaboration},%
pdftitle={Search for dark matter and large extra dimensions in monojet events in pp collisions at sqrt(s)= 7 TeV},%
pdfsubject={CMS},%
pdfkeywords={CMS, physics, dark matter, ADD, extradimensions, unparticle, monojet}}

\maketitle
\section{Introduction}
A search for new physics has been made based on events containing a jet and an imbalance in transverse momentum (\MET) in a data sample corresponding to an integrated luminosity of 5.0\fbinv. The data were collected with the Compact Muon Solenoid (CMS) detector in $\Pp\Pp$ collisions provided by the Large Hadron Collider (LHC) at a center-of-mass energy of 7\TeV.
This search is sensitive to beyond the standard model particles that do not interact in the CMS detector and whose presence
can thus only be inferred by the observation of $\MET$. The signature has been proposed as a discovery signal for many new physics scenarios.
In this paper, we use this signature to constrain
the pair production of dark matter particles~\cite{bib:TMTait,bib:RoniHarnik} and large extra dimensions in the
framework of the model proposed by Arkani--Hamed, Dimopoulos, and Dvali (ADD) \cite{bib:ADD1,ADDPRD,Antoniadis,ADDGiudice,ADDPeskin}.
The primary backgrounds to this signature arise from the production of $\cPZ$+jet and $\PW$+jet events.

Dark matter (DM) is required to accommodate numerous astrophysical measurements, such as the rotational speed of galaxies and
gravitational lensing~\cite{{DarkMatterReview},{WMAP},{DMGeneral}}.
One of the best candidates for dark matter is a stable weakly interacting massive particle.
These particles may be pair-produced at the LHC provided their mass is less than half the parton center-of-mass energy, $\sqrt{\hat{s}}$.
When accompanied by a jet from initial state radiation (ISR), DM events will have the signature of a jet plus missing transverse momentum.
The interaction between the dark matter particle (\DM) and standard model (SM) particles can be assumed to be mediated by a heavy particle
such that it can be treated as a contact interaction, characterized by a scale $\Lambda= M/\sqrt{g_{\chi} g_{q}}$ where $M$ is the mass of the
mediator, $g_{\DM}$ and  $g_{q}$ are its coupling to \DM and to quarks, respectively~\cite{bib:RoniHarnik}.
In this paper, results for the vector and axial-vector interactions between \DM and quarks are presented,
assuming \DM is a Dirac fermion.
The vector interaction can be related to spin-independent DM-nucleon  whereas axial-vector interaction can be converted to
spin-dependent DM-nucleon interactions.
The results are not greatly altered if the DM particle is a Majorana fermion,
although the vector interactions are not present in this case~\cite{bib:RoniHarnik}.

Results from previous collider searches in the monojet plus \MET channel~\cite{{bib:CMS_EXO11003},{bib:ATLASMonoJet}}
have been used to set limits on the dark matter-nucleon scattering cross section ($\sigma_{\DM N}$)~\cite{bib:RoniHarnik,RoniLHC}.
Limits on $\sigma_{\DM N}$ have also been determined
by the CMS Collaboration in the monophoton plus \MET channel~\cite{CMSMonophotons},
and by the CDF Collaboration in the monojet channel~\cite{Aaltonen:2012jb}.
 Dark matter particle production results from colliders can be compared with results from searches for dark matter-nucleon scattering
(direct detection)~\cite{bib:XENON100,bib:COGENT,bib:COUPP,bib:CDMSII2011,bib:CDMSII2010,bib:PICASSO,SIMPLE1}
and from searches for dark matter annihilation (indirect detection)~\cite{{IceCube:2011aj},{SUPERK}}.
Indirect detection experiments assume that the DM particle is a  Majorana fermion.

The ADD model accommodates the large difference between
the electroweak and Planck scales by introducing a number $\delta$ of
extra spatial dimensions, which in the simplest scenario are
compactified over a multidimensional torus of common radius $R$.
In this framework, the SM particles and gauge interactions are confined to the ordinary $3 + 1$ space-time dimensions,
 whereas gravity is free to propagate through the
entire multidimensional space. The strength of the gravitational force in $3 + 1$ dimensions is effectively diluted.
The fundamental scale  $\MD$ of this 4+$\delta$-dimensional theory is related to the apparent four-dimensional Planck scale $\Mpl$
according to $\Mpl^2\approx \MD^{\delta+2} R^{\delta}$.
The production of gravitons is expected to be greatly enhanced by the increased phase space available in the extra dimensions.
Once produced, the graviton escapes undetected into extra dimensions and its presence must be inferred from \MET.
Searches for large extra dimensions in monojet or monophoton channels were performed previously
\cite{bib:ALEPH,bib:OPAL,bib:DELPHI,bib:L3,bib:CDFMonoPhoton,bib:D0MonoPhoton,D0MonoJets,bib:CMS_EXO11003,bib:ATLASMonoJet},
and no evidence of new physics was observed.
The current lower limits on $\MD$ range from 3.67\TeVcc for $\delta=2$ to
2.25\TeVcc for $\delta=6$~\cite{bib:CMS_EXO11003}.

This paper is organized as follows.
Section 2 contains a brief description of the CMS detector and event reconstruction, and this is followed by a description of signal and
SM event simulation in Section 3.
In Section~4 we present the event selection.
The determination of dominant backgrounds from data is described in Section 5 and the results are given in Section~6.
The conclusions are summarized in Section~7.

\section{The CMS detector and event reconstruction}
CMS uses a right-handed coordinate system in which the $z$ axis
points in the anticlockwise beam direction, the $x$ axis points towards the center of the LHC ring,
and the $y$ axis points up, perpendicular to the plane of the LHC ring.
The azimuthal angle $\phi$ is measured in the $x$-$y$ plane, and the polar angle $\theta$ is measured with respect to the $z$ axis.
A particle with energy $E$ and momentum $\vec{p}$ is characterized by
transverse momentum $\pt = |\vec{p}|\,\sin{\theta}$,
and pseudorapidity $\eta = -\ln\left[ \tan(\theta/2) \right]$.

The CMS superconducting solenoid, 12.5\unit{m} long with an internal diameter of 6\unit{m}, provides a uniform magnetic field of
3.8\unit{T}. The inner tracking system is composed of a pixel detector with
three barrel layers at radii between 4.4 and 10.2\unit{cm}
and a silicon strip tracker with 10 barrel detection layers extending
outwards to a radius of 1.1\unit{m}. This system is complemented by two
endcaps, extending the acceptance up to $|\eta|=2.5$.
The momentum resolution for reconstructed tracks
in the central region is about 1\% at $\pt$ = 100\GeVc.
The calorimeters inside the magnet coil consist of a
lead tungstate crystal electromagnetic calorimeter (ECAL)  and a
brass-scintillator hadron calorimeter (HCAL) with coverage up
to $|\eta|=3$.
The quartz/steel forward hadron calorimeters extend the calorimetry coverage up to $|\eta|=5$.
The HCAL has an energy resolution of about 10\% at 100\GeV for charged pions.
Muons are measured up to $|\eta|<2.4$ in gas-ionization detectors embedded in the flux-return yoke of the magnet.
A full description of the CMS detector can be found in Ref.~\cite{bib:CMS_TDR}.

Particles in an event are individually identified using a particle-flow
reconstruction~\cite{bib:ANA_PF}. This algorithm reconstructs each particle produced in a collision by combining information from the tracker,
the calorimeters, and the muon system, and identifies them as either charged hadrons, neutral hadrons, photons, muons, or electrons.
These particles are used as inputs to the anti-$\kt$ algorithm~\cite{bib:ANA_AK} with a distance parameter of 0.5.
Jet energies are corrected to particle level with $\pt$- and $\eta$-dependent correction factors.
These corrections are derived
from Monte Carlo (MC) simulation and, for data events, are supplemented by a correction,
derived by measuring the \pt balance in dijet
events from collision data~\cite{JETJINST}.
The \MET in this analysis is defined as the magnitude of the vector sum of the transverse momentum
of all particles reconstructed in the event excluding muons.
This definition allows the use of a control sample of \Zmumu events for estimating the \Znunu background.

Muons are reconstructed by finding compatible track segments in the silicon tracker
and the muon detectors~\cite{bib:ANA_muons} and
are required to be within $|\eta|<2.1.$
Electron candidates are reconstructed starting from a cluster of energy deposits in the ECAL
that is then matched to the energy associated with a track in the silicon tracker.
Electron candidates are required to have $|\eta|<1.44$ or
$1.56<|\eta|<2.5$ to avoid poorly instrumented regions. %
Muon and electron candidates are required to originate within 2\,mm of the beam axis in the
transverse plane.
Muons (electrons) are also required to be spatially separated from jets by at
least $\Delta R=\sqrt{(\Delta \eta)^2+(\Delta \phi)^2}=0.3$, where $\Delta\eta$ and $\Delta\phi$ are differences between the muon (electron) and jet directions
in pseudorapidity and azimuthal angle, respectively.
A relative isolation parameter is defined as the sum of the \pt\ of the charged hadrons, neutral hadrons, and photon contributions computed in a
cone of radius 0.3 around the lepton direction, divided by the lepton \pt.
Lepton candidates with relative isolation values below 0.2 are considered isolated.

\section{Monte Carlo event generation}
The DM signal samples, consisting of $\chi\bar{\chi}$ pairs associated with one parton, are produced using the leading order (LO) matrix element event
generator \MADGRAPH~\cite{bib:GEN_Mg} interfaced with \PYTHIA 6.42~\cite{bib:GEN_Py6} with tune Z2~\cite{bib:GEN_Py6_D6T}
for parton showering and hadronization. Dark matter particles masses $M_\chi$ =0.1, 1, 10, 200, 300, 400, 700, and 1000\GeVcc are generated for
both vector and axial-vector interactions. In addition, the \pt of the associated parton is required to be greater than 80\GeVc.
The parton showering program generates partons in a phase space that overlaps with the phase space
of the partons generated by the matrix element calculator.
Double-counting by the matrix element calculation and parton showering is resolved by using the {\sc mlm} matching prescription~\cite{bib:ALPGEN},
as implemented in~\cite{bib:GEN_Mg}. The CTEQ~6L1~\cite{bib:SYST_CTEQ6M} parton distribution functions (PDF) are used.

The events for the ADD model are
generated with \PYTHIA 8.130~\cite{bib:GEN_PY8,bib:GEN_Ask}, using tune 4C~\cite{TuneFourC} and the CTEQ~6.6M~\cite{bib:SYST_CTEQ6M} PDFs.
This model is an effective theory and holds only for energies well below $\MD$.
For a parton center-of-mass energy $\sqrt{\hat{s}}>\MD$, the simulated cross sections of the graviton are suppressed by a factor $\MD^4/\hat{s}^2$~\cite{bib:GEN_Ask}.
Because the $\sqrt{\hat{s}}$ values for the data are smaller than the current limits on $\MD$, the results are not affected by this suppression.
The next-to-leading-order (NLO) QCD corrections
to direct graviton production in the ADD model are sizable and depend
on the \pt of the recoiling parton~\cite{bib:Grav_NLO}.
As a simplifying assumption, we use  $K$-factors ($\sigma_{\textrm{NLO}}/\sigma_{\textrm{LO}}$) corresponding to a fixed graviton \pt of several hundred \GeVc;
the values are 1.5 for $\delta$ = 2, 3 and 1.4 for $\delta$ = 4, 5, and 6.

The $\cPZ$+jets, $\PW$+jets, \ttbar, and single-top event samples are produced using \MADGRAPH interfaced with
\PYTHIA 6.42, using tune Z2 and the CTEQ~6L1 PDFs. They are normalized to NLO cross sections~\cite{Melnikov:2006kv}. The QCD multijet
sample is generated with \PYTHIA 6.42, using tune Z2 and CTEQ~6L1 PDFs and \PYTHIA LO cross sections are used.
All the generated signal and background events are passed through a
\GEANTfour~\cite{Agostinelli2003250}
simulation of the CMS detector.

\section{Event selection}
\label{sec:ANA}
The data used in this analysis were recorded by a trigger that required an event to have a jet with $\pt > 80$\GeVc and
$\MET > 80$ or 95\GeVc as measured online by the trigger system. The threshold of  80 (95) \GeVc was used to collect 4.2 (0.87)\fbinv of data.

Events are required to have at least one primary vertex~\cite{bib:ANA_Tk} reconstructed
within a $\pm 24$\unit{cm} window along the beam axis around the detector center, and a transverse distance from the beam axis of less than 2\cm.
Signals in the calorimeter that are not associated with pp interactions are identified based either on energy sharing
between neighboring channels or timing requirements and are excluded from further
reconstruction~\cite{bib:ANA_JME-10-004}.

To suppress the remaining instrumental and beam-related backgrounds, events are rejected if less than 20\% of the energy of the highest \pt jet
is carried by charged hadrons or more than 70\% of this energy is carried by either neutral hadrons or photons. Events are also rejected if
more than 70\% of the \pt of the second highest \pt jet is carried by neutral hadrons.
Such spurious jets primarily arise from instrumental noise, where the energy deposition is limited to one sub-detector. Jets resulting from
energy deposition by beam halo or cosmic-ray muons do not have associated tracks and are also rejected by these selections.
All events passing these selection requirements and with $\MET > 500$\GeVc were visually inspected and found to be consistent with pp collision events.
The application of these data cleanup requirements would reject approximately 2\% of the dark matter signal and 3\% of the ADD signal.

The signal sample is selected by requiring $\MET>200\GeVc$ and
the jet with the highest transverse momentum ($\,\mathrm{j}_1$) to have $\pt(\,\mathrm{j}_1)>110\GeVc$ and
$|\eta(\,\mathrm{j}_1)|<2.4$.
The triggers used to collect these data are fully efficient for events passing these selection cuts.
Events with more than two jets with $\pt$ above 30\GeVc are discarded.
As signal events typically contain jets from initial- or final-state radiation, a second jet ($\,{\rm j}_2$) is allowed, 
provided $\Delta\phi(\,\mathrm{j}_1,\mathrm{j}_2)<2.5$\rad.
This angular requirement suppresses QCD dijet events.
To reduce background from Z and W production and top-quark decays, events with isolated muons or electrons with
$\pt>10\GeVc$ are rejected. Events with an isolated track with $\pt > 10\GeVc$ are also removed as they
come primarily from $\Pgt$-lepton decays.
A track is considered isolated if the scalar sum of the transverse momentum of all tracks with $\pt > 1\GeVc$ in the annulus of $0.02<\DR<0.3$ around its direction is less than 1\% of its \pt.
Table 1 lists the numbers of data and SM background events at each
step of the analysis. Efficiencies for representative dark matter and
ADD models relative to the event yield passing $\MET>200\GeVc$
selection are also shown.
The dominant background is \ZnunuJets and the next largest source of background is \WJets.
The event yields for $\MET > 250,\, 300,\, 350$, and 400\GeVc are also shown.
A study of the $\MET$ requirement using the signal samples showed that $\MET > 350$\GeVc is the optimal value for both the dark matter and ADD models searches.

\begin{figure}[bt]
  \begin{center}
  {\includegraphics[scale=0.39]{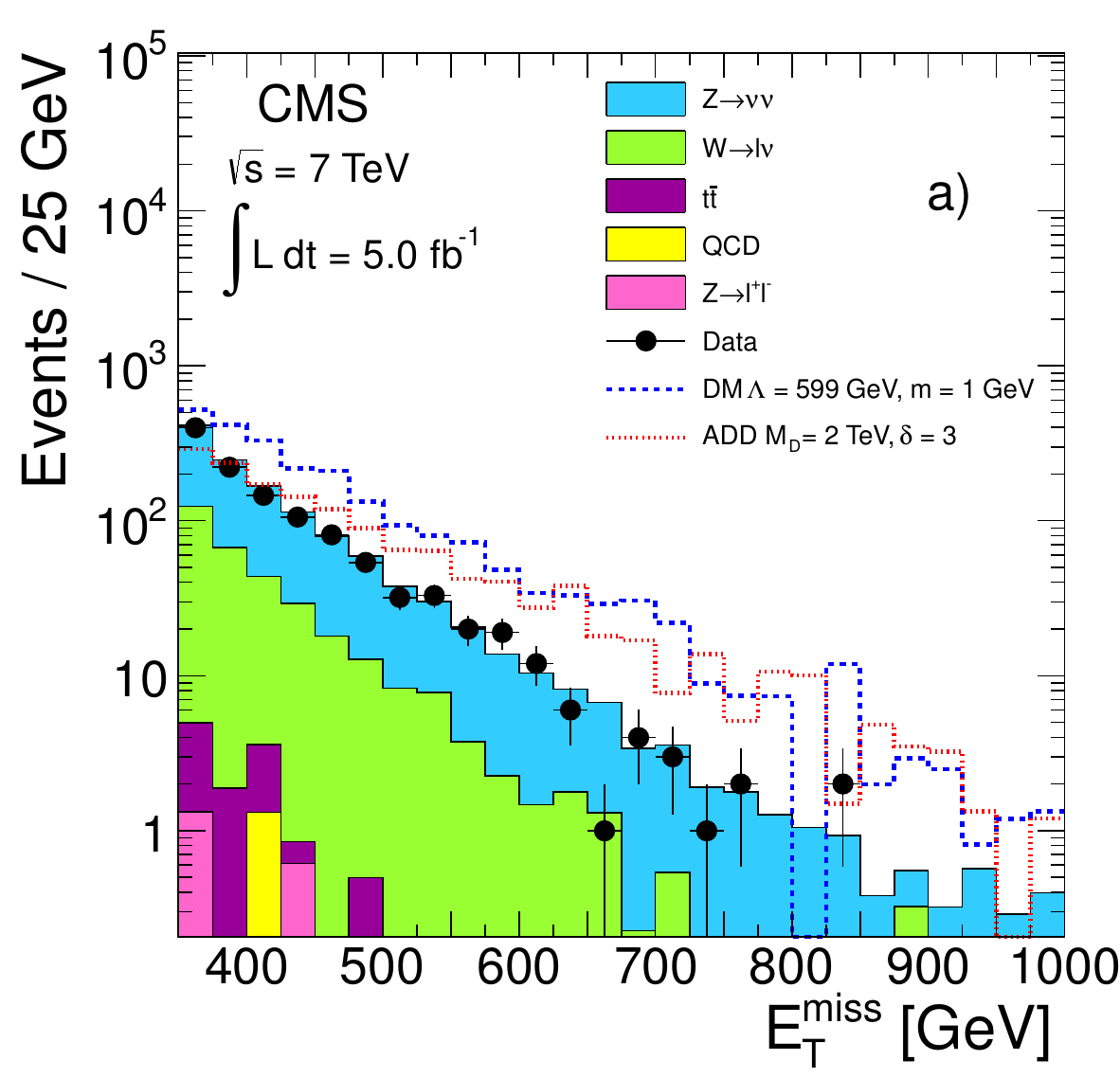}
  \includegraphics[scale=0.39]{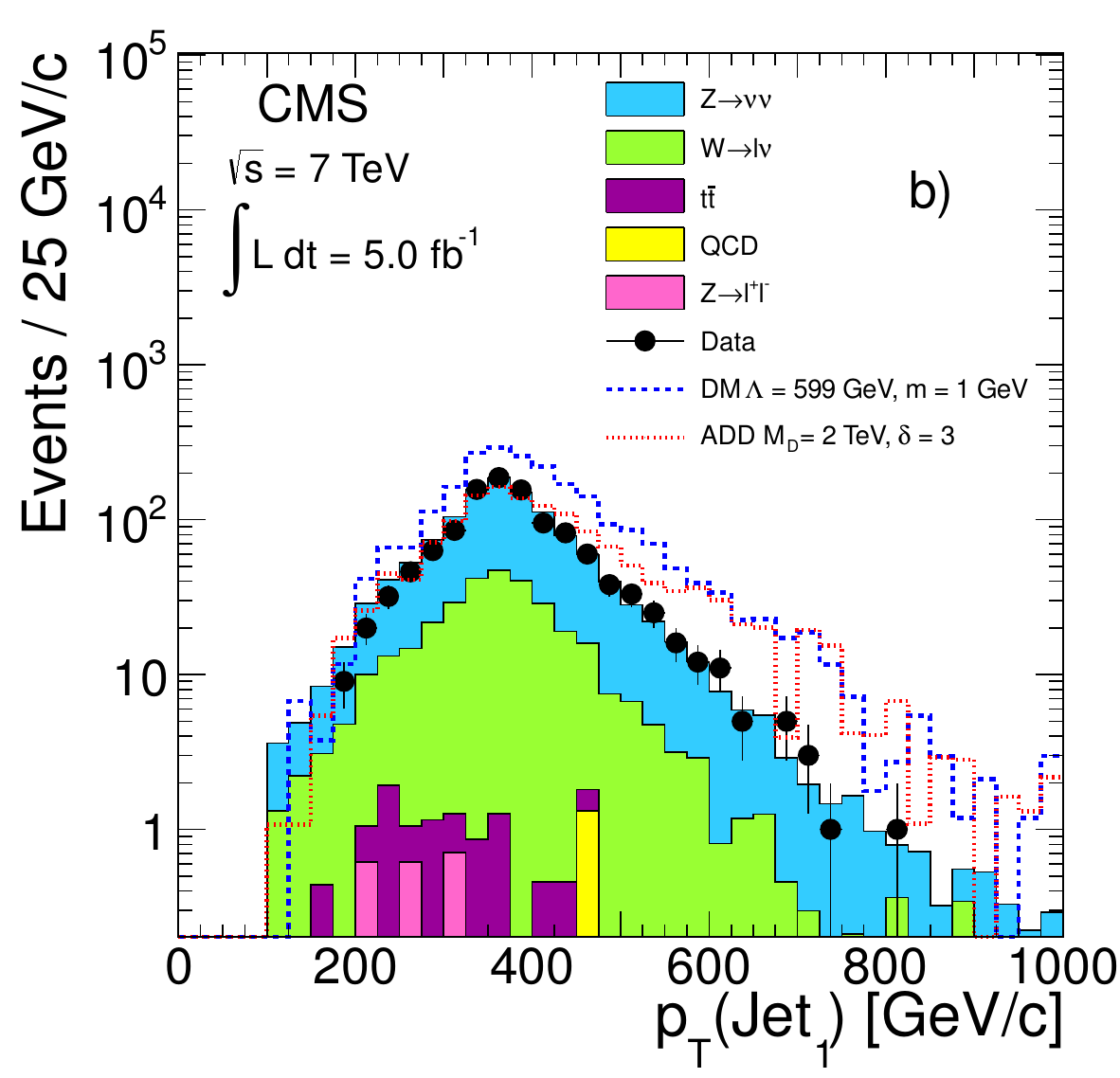}}
  \caption{The distribution of (a) \MET and (b) $\pt(\,\mathrm{j}_1)$ for data (black full points with error bars) and simulation (histograms) for $\MET > 350$ \GeVc after the full event selection criteria are applied.
The $\cPZ(\cPgn\cPgn)$+jets and $\cPW$+jets backgrounds are normalized to their estimates from data. An example of a dark matter
signal (for axial-vector couplings and $M_{\chi} = 1$\GeVcc)
is shown as a dashed blue histogram and an ADD signal (with $\MD=2\TeV$, $\delta=3$) is shown as a dotted red histogram.
\label{fig:ANA_Jet1_plot}}
\end{center}
\end{figure}
The \MET and $\pt(\,\mathrm{j}_1)$ distributions are shown in Fig.~\ref{fig:ANA_Jet1_plot}, where the \ZnunuJets and $\cPW$+jets backgrounds are normalized to
the rate determined from data (Section 5)
and other backgrounds are normalized to the integrated luminosity.

\begin{landscape}
\begin{table}[p]
\begin{center}
\topcaption{Event yields at different stages of the event selection
 for  (a)  various SM processes from simulation,  (b)  sum of all
 SM processes, and the  data, corresponding to an integrated luminosity of 5.0\fbinv. Only statistical uncertainties are shown, which in most cases are
smaller than the associated systematic uncertainties.
Lepton removal eliminates events with isolated electrons, muons, or tracks with $\pt \!>10\GeVc$.
Efficiencies for representative dark matter and ADD models relative to the event yield passing $\MET>200\GeVc$ selection are also given.}
\label{tab:SEL_TabDataMC}
\begin{tabular}{lcccccc}
\multicolumn{7}{c}{ (a) } \\ \hline
Requirement
& $\cPZ(\cPgn\cPgn)$
& $W(\Pe\nu,\mu\nu,\tau\nu)$
& $\cPZ(\ell\ell)$
& \ttbar & Single t
& \multicolumn{1}{c}{QCD}
\\
&+jets &+jets &+jets & & & Multijet  \\ \hline
\MET $>$200\GeVc                            & $ (324\pm 1)\times  10^2    $&  $(591\pm 1)\times 10^2 $ & $ 5255\pm 47             $ & $  (133\pm 1)\times 10^2  $ & $  1165\pm 7   $ & $ (160 \pm 1)\times 10^2    $\\
$\pt(\,\mathrm{j}_1)>110\GeVc$                   & $ (302\pm 1)\times  10^2    $&  $(557\pm 1)\times 10^2 $ & $ 4908\pm 45             $ & $  (119\pm 1)\times 10^2  $ & $  1035\pm 6   $ & $ (158 \pm 1)\times 10^2    $\\
${{N}_\text{Jet}} (\pt>30\GeVc)\le 2$       & $ (227\pm 1)\times  10^2    $&  $(397\pm 1)\times 10^2 $ & $ 3453\pm 38             $ & $  1587\pm 20             $ & $   274\pm 3   $ & $  5296\pm 14               $\\
$\Delta\phi(\,{\rm jet}_1, {\rm jet}_2)<$2.5   & $ (211\pm 1)\times  10^2    $&  $(354\pm 1)\times 10^2 $ & $ 3139\pm 36             $ & $  1344\pm 19             $ & $   237\pm 3   $ & $  62\pm 5                  $\\
Lepton Removal                               & $ (198\pm 1)\times  10^2    $&  $ (97\pm 1)\times 10^2 $ & $   81\pm 6              $ & $   214\pm  7            $ & $    35\pm 1    $ & $   2\pm 1                  $\\
\MET $>$250\GeVc                            & $   7306\pm 23              $&  $  2951\pm 19          $ & $   22\pm 3              $ & $    70\pm  4            $ & $    10\pm 0.5  $ & $   2\pm 1                  $\\
\MET $>$300\GeVc                            & $   2932\pm 14              $&  $  967\pm 11           $ & $    6\pm 1.7            $ & $    23\pm  2            $ & $     3\pm 0.2  $ & $   1\pm 0.7                $\\
\MET $>$350\GeVc                            & $   1308\pm 9               $&  $  362\pm 7            $ & $    2\pm 0.9            $ & $     9\pm  1.6          $ & $     1\pm 0.2  $ & $   1\pm 0.7                $\\
\MET $>$400\GeVc                            & $    628\pm 7               $&  $  148\pm 4            $ & $    1\pm 0.4            $ & $     3\pm  0.8          $ & $   0.4\pm 0.1  $ & $   1\pm 0.7                $\\
\hline
\end{tabular}
\vspace{5mm}\\
\begin{tabular}{lcccc}
\multicolumn{5}{c}{(b)} \\ \hline
  \multicolumn{1}{c}{Requirement}
& \multicolumn{1}{c}{Total}
& \multicolumn{1}{c}{Data}
& \multicolumn{1}{c}{ DM (\%)}
& \multicolumn{1}{c}{ADD (\%)}
\\  

& Simulated SM
&
& $M_{\DM}=1$\GeVcc
& \MD=2\TeVcc,$\delta=3$ \\ \hline
\MET $>$200 \GeVc                             & $ 1273  \times 10^2    $ & $ 1045\times   10^2 $ &  $  100.0          $   &  $   100.0         $   \\
$\pt(\,\mathrm{j}_1)>110\GeVc$                 & $ 1195  \times 10^2    $ & $ 1007\times   10^2 $ &  $   95.2 \pm 0.71 $   &  $   92.8\pm1.03   $   \\
${{N}_\text{Jet}} (\pt>30\GeVc)\le 2$      & $ 730   \times 10^2    $ & $ 624\times    10^2 $ &  $   69.8 \pm 0.60 $   &  $   61.5\pm0.84   $   \\
$\Delta\phi(\,\mathrm{j}_1, \mathrm{j}_2)<$2.5     & $ 613   \times 10^2    $ & $ 54\times 10^2     $ &  $   66.8 \pm 0.59 $   &  $   58.0\pm0.81   $   \\
 Lepton Removal                              & $ 298   \times 10^2    $ & $ 24\times 10^2     $ &  $  63.7 \pm 0.58  $   &  $    54.8\pm0.79  $   \\
\MET $>$250 \GeVc                             & $ 104   \times 10^2    $ & $ 76\times 10^2     $ &  $  34.1 \pm 0.42  $   &  $    32.5\pm0.61  $   \\
\MET $>$300 \GeVc                             & $   3932               $ & $  2774             $ &  $  18.9 \pm 0.31  $   &  $    19.9\pm0.48  $   \\
\MET $>$350 \GeVc                             & $   1683               $ & $  1142             $ &  $  10.8 \pm 0.24  $   &  $    12.5\pm0.38  $   \\
\MET $>$400 \GeVc                             & $    782               $ & $   522             $ &  $   6.4 \pm 0.18  $   &  $     7.9\pm0.30  $   \\
\hline
\end{tabular}
\end{center}
\end{table}
\end{landscape}

\section{Background estimate from data}
Table~\ref{tab:SEL_TabDataMC} shows that the SM backgrounds remaining after the full event selection are dominated
by the following processes:  \ZJets with the \Z boson decaying into a pair of neutrinos and \WJets with the \W boson decaying leptonically.
These backgrounds are estimated from data utilizing a control sample of $\mu$+jet events, where \Zmumu events are used to estimate
the $\cPZ(\cPgn\cPgn)$ background and $\cPW(\mu\cPgn)$ events are used to estimate the remaining \WJets background.
The control sample is derived from the same set of triggers as those used to collect the signal sample by applying the
full event selection criteria except for the vetoes on electrons,
muons, and isolated tracks. One or more isolated muons with $\pt>20$\GeVc and $|\eta| < 2.1$ are required.

\begin{figure}[!Hhtb]
  \begin{center}
\vspace{5mm}
  \includegraphics[scale=0.35]{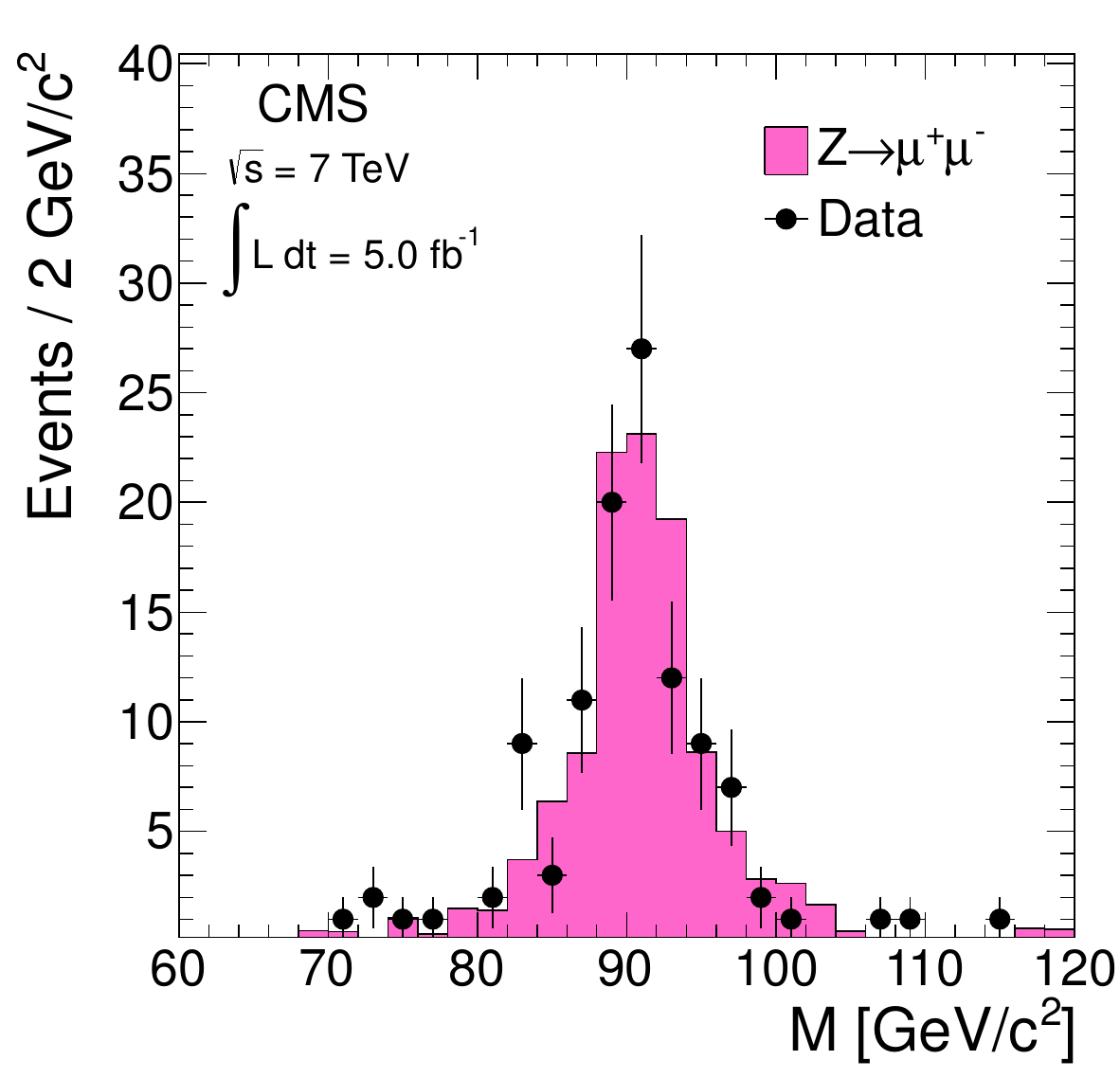}
  \caption{The dimuon invariant mass distribution in the dimuon control sample in data (black full points with error bars) and simulation (histogram) for $60 < M_{\mu\mu} < 120$\GeVcc.
   The MC prediction has been normalized to the data yields. There is no significant non-\Z background.}
  \label{fig:BKGR_Z_mass}
  \end{center}
\end{figure}

A sample of \Zmumu events is selected by requiring two isolated muons with opposite-sign charges and a dimuon invariant mass between 60 and 120\GeVcc.
The observed yield is 111 events, which should be compared with a mean expected yield of 136$\pm$8 events, where the uncertainty is only statistical.
The dimuon invariant mass distributions, both for the data control sample and for the simulation, are shown in Fig.~\ref{fig:BKGR_Z_mass}.

The production of a \Z boson in association with jets and its subsequent decay into neutrinos has characteristics that are similar to those in the production of \ZJets where
the \Z decays to muons. Thus by treating the pair of muons as a pair of neutrinos, the topology of the \Znunu process is reproduced.
The number of $\cPZ(\cPgn\cPgn)$ events can then be predicted using:
\begin{equation}
N(\Znunu) = \frac{N^{{\rm obs}} - N^{{\rm bgd}}}{A\times\epsilon}\cdot R\left(\frac{\Znunu}{\Zmumu}\right)
\end{equation}
where $N^{\rm obs}$ is the number of dimuon events observed, $N^{{\rm bgd}}$ is the estimated number of background events contributing to the dimuon sample, $A$ is the geometric and kinematic acceptance
 of the detector and the $Z$ mass window, $\epsilon$ is the selection efficiency for the event, and $R$ is the ratio of branching fractions for the $Z$ decay to a pair of
 neutrinos and to a pair of muons.

The acceptance A is defined as the fraction of all simulated events that pass all signal selection requirements except muon and track veto and have two muons with $\pt > 20$\GeVc and $|\eta|< 2.1$ and with an invariant mass within the Z mass window.  The selection efficiency $\epsilon$ is defined as the fraction of the events passing acceptance cuts that have two reconstructed muons with $\pt> 20$\GeVc and $|\eta| < 2.1$ and with an invariant mass within the Z mass window.
This efficiency is estimated from simulation.
The muon selection efficiency, both in the data and the simulation, is determined in the dimuon 
events with one of the muons passing tight selection criteria (tag)
and with an invariant mass in the \Z boson mass window.
The efficiency of the second muon (probe), assumed 
to be a muon originating from the decay of the \Z boson after background subtraction, is determined for the selection requirements used in 
this analysis. Details of this ``tag-and-probe'' method can be found  in Ref.~\cite{bib:tagprobe}. The efficiencies in the data and the simulation are consistent.
The stability of this agreement is measured by varying the muon kinematics and  
the largest difference between the efficiencies in the data and the simulation is assigned as the uncertainty on the muon selection. This translates into 2\% systematic 
uncertainty on $\epsilon$.
The ratio of the branching fractions $R$ is $5.942\pm 0.019$~\cite{bib:BKG_PDG}.
Some of the \ZnunuJets events would be rejected by the track isolation requirement, and the background is multiplied by a factor of 0.94 to account for this effect.
The scaling factor is obtained from simulation.

The final prediction for the number of $\Znunu$ events is 900 $\pm$ 94 for $\MET > 350$\GeVc, where the uncertainty includes statistical and systematic contributions.
The sources of this uncertainty are: (i) the statistical uncertainties on the number of \Zmumu events in the data and simulation,
(ii) uncertainties
 on the acceptance from PDF uncertainties, evaluated based on the PDF4LHC~\cite{bib:PDF4LHC} recommendations,
and (iii) the uncertainty in the selection efficiency $\epsilon$ as determined from the difference in measured efficiencies in data and MC simulation.
Table~\ref{tab:zinv_sys} summarizes the systematic uncertainties.

\begin{table*}[!Hhtb]  
\begin{center}
\caption{Sources of systematic uncertainty and their fractional contributions to the total uncertainty on the \Znunu background.}
\label{tab:zinv_sys}
\begin{tabular}{l|c} \hline
Source of Uncertainty              & Size (\%)  \\ \hline
Size of control sample ($N_{\rm obs}$) & 9.5\\
Geometric and kinematic acceptance (A)        &  3.7\\
Muon selection efficiency ($\epsilon$) &   2.1 \\
Track isolation selection efficiency & 3.6 \\
Ratio of branching fractions (R)             &  0.3 \\ \hline
Total                 &  11.0 \\ \hline

\end{tabular}
\end{center}
\end{table*}

The second largest background arises from \WJets events that are not removed by the  lepton veto cut. These events can come from events in which
the lepton (electron or muon) is either not identified, not isolated, or out of the acceptance region, or events in which a $\tau$ decays hadronically.
The events where the lepton is `lost' are estimated from the $\Wmunu+$jets control sample.

\begin{figure}[!Hhtb]
\begin{center}
\vspace{5mm}
  \includegraphics[scale=0.35]{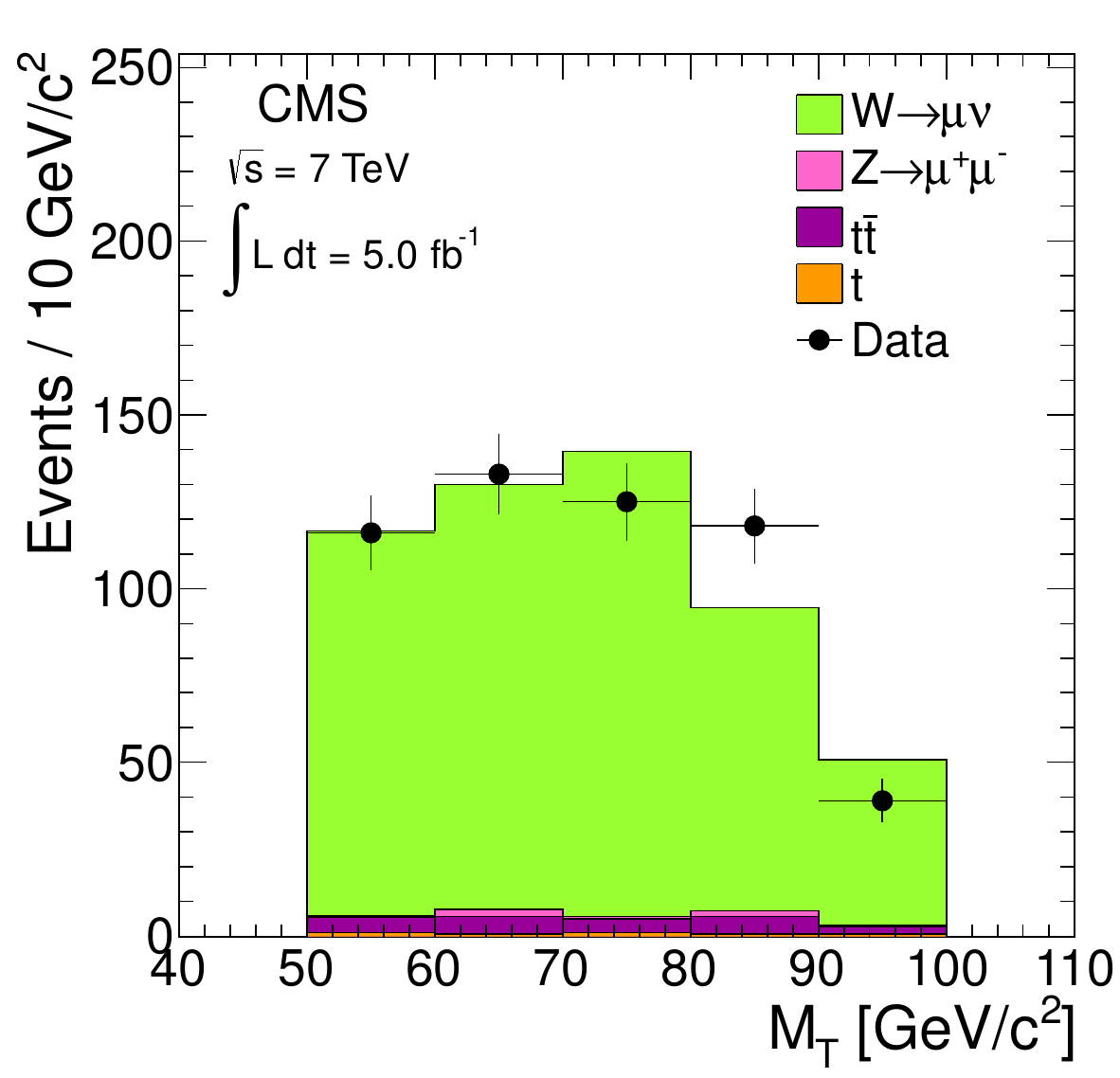}
  \caption{The transverse mass distribution $M_{\rm T}$ in the single muon data control sample and MC predictions for \Wmunu, \ttbar, \Zmumu, and single top-quark production.
   The MC predictions have been normalized to the data yields. Data are dominated by \Wmunu events.}
  \label{fig:BKGR_W_mt}
  \end{center}
\end{figure}

A \Wmunu sample is selected by requiring an isolated muon with $\pt > 20$\GeVc and $|\eta| < 2.1$ and the transverse mass $M_{\rm T}$ to be
between 50 and 100\GeVcc.  The transverse mass is defined
as $M_{\rm T}=\sqrt{2\pt^{\mu}\MET\left(1-\cos(\Delta\phi\right))}$, where $\pt^{\mu}$ is the transverse momentum of the muon and
$\Delta\phi$ is the angle between the muon \pt and the \MET vectors. The event yields obtained for the $\cPW(\mu\cPgn)$ sample
for $\MET > 350$\GeVc are shown in Table~\ref{tab:Wmuontable}, along
with the contributions from Z+jets, \ttbar, and single top-quark
events predicted by the simulation.
The observed yield of \WmunuJet candidates is 531 which can be
compared with a mean expected yield of $615.4\pm 9.3$, where uncertainty  is statistical only.
Figure~\ref{fig:BKGR_W_mt} shows the $\cPW$ transverse mass distribution for data and simulation in the $\cPW(\mu\cPgn)$ control sample.

\begin{table*}[!Hhtb]
\begin{center}
\caption{Event yields for the $\cPW(\mu\cPgn)$ from simulation including non-\W backgrounds, and from the data control sample.}
\label{tab:Wmuontable}
\begin{tabular}{cccc|cc} \hline
                           $W$+jets &  \ttbar & $Z$+jets &  Single t    & All MC & Data\\\hline
 581.5 & 23.3 & 6.4 & 4.2  & 615.4  & 531  \\ \hline
\end{tabular}
\end{center}
\end{table*}

$\cPW(\mu\cPgn)$ candidate events ($N_{\rm obs}$), after subtracting non-\W contamination ($N_{\rm bgd}$),
are corrected for the detector acceptance ($A'$) and selection efficiency ($\epsilon'$) to obtain the total number of produced events
$N_{\rm tot} = (N_{\rm obs} - N_{\rm bgd})/(A'\times\epsilon')$.
This number is subsequently weighted by the inefficiency of the selection criteria used in the definition of the lepton veto to predict the number of events that are not rejected by the
veto and thus remain in the signal sample.

The number of $\cPW(\mu\cPgn)+$jet events that are either out of the acceptance ($N_{\bar{A}}$) or are not identified or are not isolated ($N_{\bar{\epsilon}}$) can be written as:
\begin{eqnarray}
N_{\bar{A}} &=& N_{\rm tot}\times(1 - A)\\
N_{\bar{\epsilon}} &=& N_{\rm tot}\times A\times(1 - \epsilon)
\end{eqnarray}
where $A$ is the acceptance, and $\epsilon$ is the selection efficiency of the muon selection used in the lepton veto.
The total background from events where the muon is `lost' is then given by
\begin{equation}
N_{{\rm lost}\,\mu} = N_{\bar{A}} + N_{\bar{\epsilon}}.
\end{equation}

An estimate of the `lost' electron background is similarly obtained from the $\cPW(\mu\cPgn)$+jets data sample, correcting for the muon acceptance and
selection efficiency to obtain $N_{\rm tot}$. The ratio of the generated $\cPW(\mu\cPgn)$ and $\cPW(\Pe\cPgn)$ events passing the signal selection is
taken from simulation and used to obtain $N_\text{tot}$ for electrons. The same procedure is then applied to obtain the number of events where the
electron is either not reconstructed or not isolated or out of the acceptance.

The detector acceptance for both muons and electrons is obtained from simulation. The selection efficiency is similarly obtained from simulation but with
 an assigned systematic uncertainty to cover the largest difference in the efficiency measured in data and simulation
with the tag-and-probe method. 

There is a remaining component of the $\PW$+jets background from events where the $\PW$ decays to a $\tau$ lepton and the $\tau$ decays
hadronically, and this is estimated from simulation.
This estimate is corrected using a normalization factor obtained from the ratio of $\cPW(\mu\cPgn)$ events in data and simulation.
The estimated \WJets background is corrected to account for the fraction of events that would be rejected by the track isolation veto.
This correction factor is obtained from simulation and found to be 19\%.

The total prediction for the number of \WJets events is $ 312 \pm 35$ for $\MET > 350$\GeVc, where the uncertainty includes both statistical and systematic contributions.
The sources of this uncertainty are:
(i) the uncertainties on the number of single-muon events in the data and simulation samples,
(ii) a conservative (100\%) uncertainty on the non-\W contamination obtained from simulation,
(iii) uncertainties on the acceptance from PDFs, and
(iv) the uncertainty in the selection efficiency $\epsilon$ as determined from the difference in measured efficiency between data and simulation.
Table~\ref{tab:W_jets_sys} summarizes the systematic uncertainties in the \WJets background.

\begin{table*}[!Hhtb]  
\begin{center}
\topcaption{Sources of systematic uncertainty and their contribution to the total uncertainty on the $\PW$+jets background.}
\label{tab:W_jets_sys}
                \begin{tabular}{l|c} \hline
Source of Uncertainty                          & Size (\%) \\ \hline
Size of control sample ($N_\text{obs}$)  &  2.9 \\
Background ($N_\text{bgd}$)  & 3.9 \\
Isolated track efficiency & 2.1 \\
Kinematic and geometrical acceptance (A)  & 7.7 \\
Selection efficiency ($\epsilon$)  & 6.8 \\ \hline
Total  & 11.6 \\ \hline
\end{tabular}
\end{center}
\end{table*}

Background contributions from QCD multijet events, \ttbar, and $\cPZ(\ell\ell)$+jets production are small
and are obtained from the simulation. A 100\% uncertainty is assigned to these background estimates.

\section{Results}
\begin{table}
\begin{center}
\topcaption{SM background predictions compared with data passing the selection requirements for various \MET thresholds,
corresponding to integrated luminosity of 5.0\fbinv. The uncertainties include both statistical and systematic terms.
In the last two rows, expected and observed  95\% confidence level upper limits on possible contributions from new physics passing the selection requirements are given.}
\label{tab:summary_bgd}
\begin{tabular}{l|cccc}
\hline
\multicolumn{1}{l|}{\MET (\GeVc) $\rightarrow$}&
$\ge 250$ &
$\ge 300$ &
$\ge 350$ &
$\ge 400$ \\
\hline
{Process} &
\multicolumn{4}{c}{Events}\\
\hline
{\ZnunuJets}        &
  5106  $\pm$ 271 &
  1908  $\pm$ 143 &
   900  $\pm$ 94  &
   433  $\pm$ 62  \\
{\WJets}      &
2632 $\pm$ 237 &
816  $\pm$ 83  &
312  $\pm$ 35  &
135  $\pm$ 17  \\
{\ttbar}        &
 69.8 $\pm$ 69.8   &
 22.6 $\pm$ 22.6   &
  8.5 $\pm$  8.5   &
  3.0 $\pm$  3.0   \\
{\ZellellJets}        &
 22.3 $\pm$22.3 &
  6.1 $\pm$ 6.1 &
  2.0 $\pm$ 2.0 &
  0.6 $\pm$ 0.6 \\
{Single t}      &
  10.2 $\pm$10.2   &
   2.7 $\pm$ 2.7   &
   1.1 $\pm$ 1.1   &
   0.4 $\pm$ 0.4   \\
{QCD Multijets}    &
 2.2  $\pm$2.2   &
 1.3  $\pm$1.3   &
 1.3  $\pm$1.3   &
 1.3  $\pm$1.3   \\ \hline
{Total SM}    &
 7842 $\pm$ 367  &
 2757 $\pm$ 167  &
 1225 $\pm$ 101  &
   573$\pm$ 65 \\
{Data}         &
 7584  &
 2774  &
 1142  &
 522   \\
\hline
{Expected upper limit non-SM}    &
779  &
325 &
200 &
118 \\
{Observed upper limit non-SM}    &
600  &
368 &
158 &
95 \\
\hline
\end{tabular}
\end{center}
\end{table}

The total number of events observed is compared with the total number of estimated background events in Table~\ref{tab:summary_bgd},
together with the breakdown of this background into separate subprocesses.
Contribution from \ZnunuJets and \WJets processes are determined from the data.
Contributions from \ttbar, \Zellell, single t, and QCD multijet processes are determined from simulation and are assumed to have 100\% uncertainty.
The number of events observed is consistent with the number of events expected from SM backgrounds.
Thus these data are used to set
limits on the production of dark matter particles and to constrain the  ADD model parameters. The CLs method~\cite{bib:STAT_RooStats,bib:BKG_PDG} is used for calculating
the upper limits on the number of signal events, and systematic uncertainties are modeled by log-normal distributions.

The important uncertainties related to signal modeling are:
\begin{enumerate}
\item The jet energy scale uncertainty, estimated by shifting the four-vectors of the jets by an $\eta$- and $\pt$-dependent factor~\cite{bib:ANA_JME-10-010}, yielding a variation of 8--11\% (8--13\%) for the dark matter (ADD) signal.
\item The noise cleaning uncertainty, obtained by assigning the full
  effect of noise cleaning as systematic uncertainty, 2\% (3\%) for dark matter (ADD) signal.
\item PDF uncertainties evaluated using the PDF4LHC~\cite{bib:PDF4LHC} prescription and resulting in a
systematic uncertainty of 1--7\% (1--4\%) for the dark matter (ADD) signal.
\item The renormalization/factorization scale uncertainty, evaluated by varying the scale up and down by a factor of two, 5\% for both dark matter and ADD signals.
\item ISR uncertainty, estimated by changing \PYTHIA parameters, yielding a variation of 15\%  for both dark matter and ADD signals.
\item Uncertainty on the pileup simulation, 3\% for both dark matter and ADD signals.
\item The limited statistics of the simulated sample yielding a variation of 2--5\% (2--4\%) on the dark matter (ADD) signal.
\end{enumerate}
The total uncertainty on the signal for the DM (ADD) models for these sources of uncertainty is 20\% (21\%).
In addition, a 2.2\% uncertainty on the integrated luminosity measurement~\cite{lumi} is included.

\begin{figure}
  \begin{center}
  \includegraphics[width=0.48\textwidth] {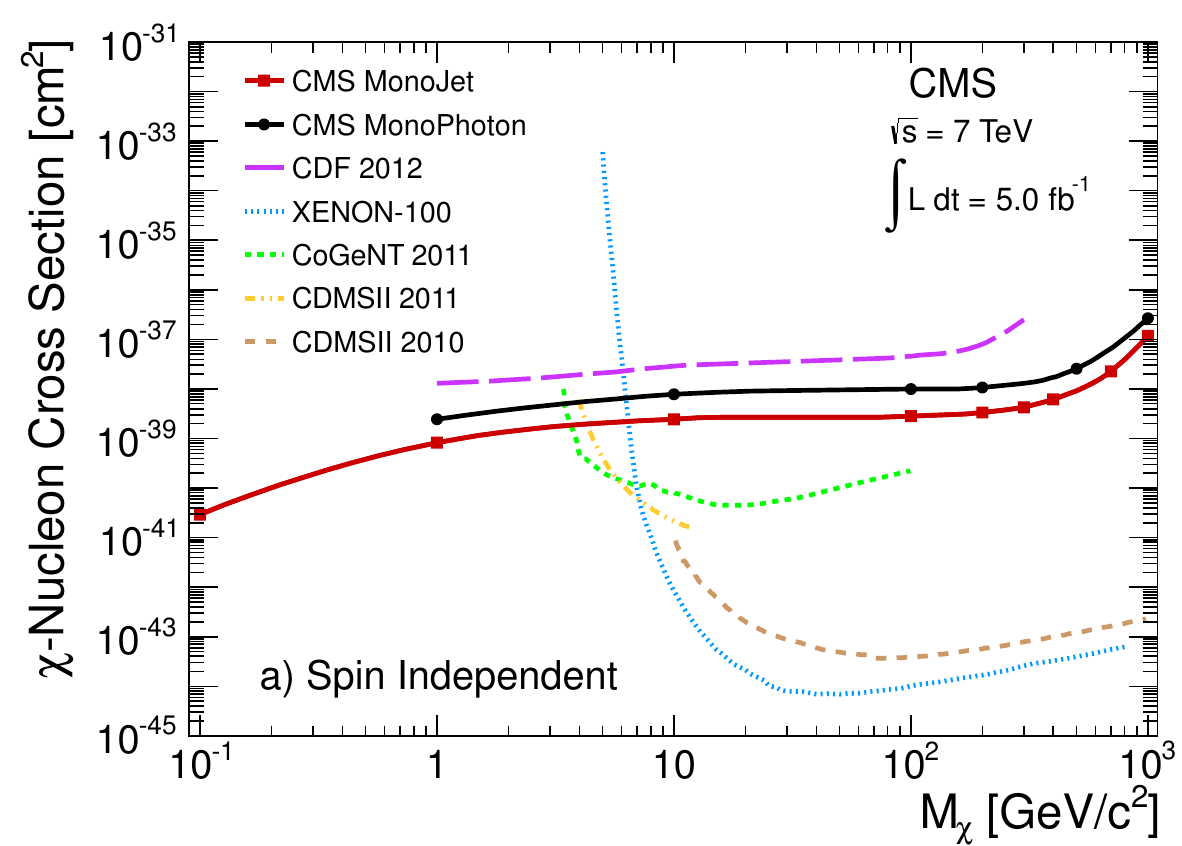}
  \includegraphics[width=0.48\textwidth] {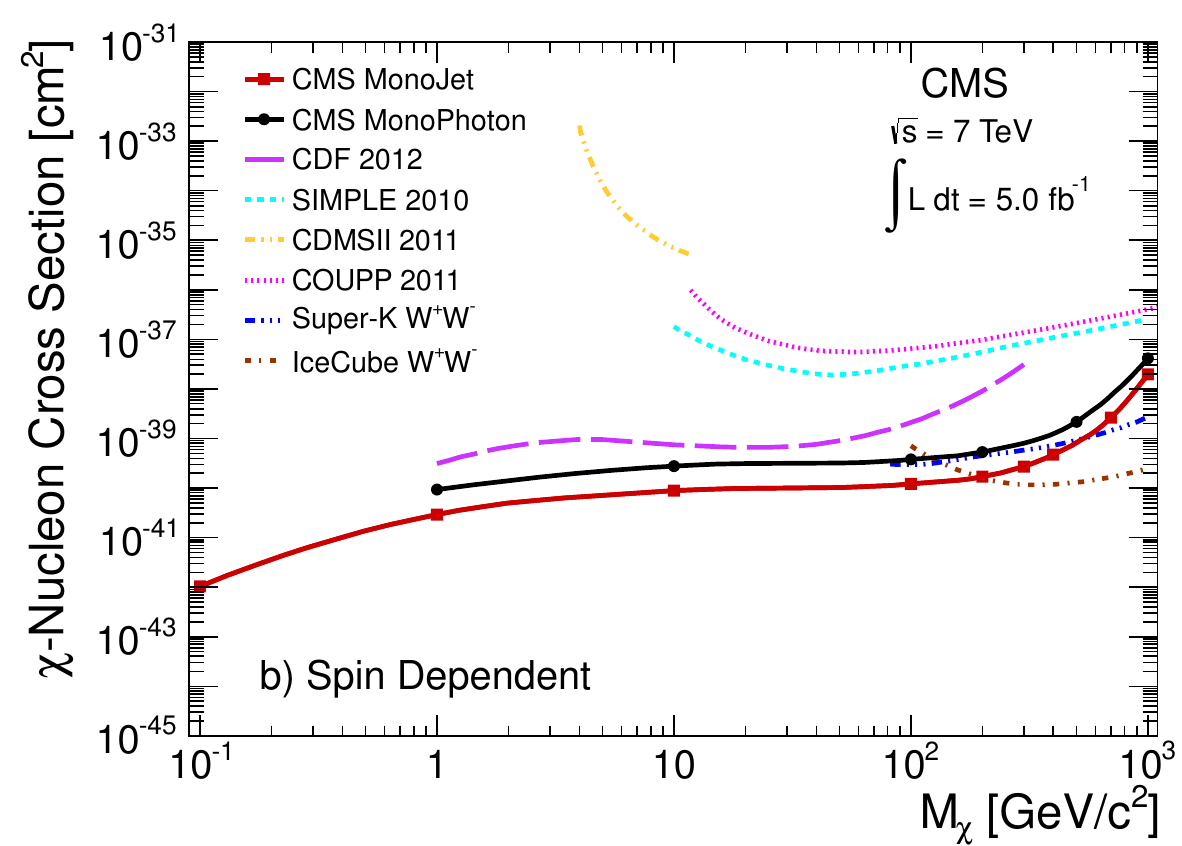}
   \caption{Comparison of the 90\% CL upper limits on the dark matter-nucleon scattering cross section versus mass of dark matter particle for the (left) spin-independent and (right) spin-dependent models with results from
CMS using monophoton signature~\cite{CMSMonophotons},
CDF~\cite{Aaltonen:2012jb},
XENON100~\cite{bib:XENON100},
CoGeNT~\cite{bib:COGENT},
COUPP\cite{bib:COUPP},
CDMS II~\cite{bib:CDMSII2011,bib:CDMSII2010},
Picasso~\cite{bib:PICASSO},
SIMPLE~\cite{SIMPLE1},
IceCube~\cite{IceCube:2011aj},
and Super-K~\cite{SUPERK} collaborations.
\label{fig:DM_limits_si}}
\end{center}
\end{figure}

For dark matter models, the observed limit on the cross section depends on the mass of the dark matter particle and the nature of its
interaction with the SM particles. The limits on the effective contact interaction scale $\Lambda$ as a function of $M_{\chi}$ can be translated into a limit on the
dark matter-nucleon scattering cross section using the reduced mass of
$\chi$-nucleon system~\cite{bib:RoniHarnik}, which can be compared with the constraints from direct and indirect detection experiments.
Figure~\ref{fig:DM_limits_si} shows the 90\% confidence level (CL)
upper limits on the dark matter-nucleon scattering cross section  as a function of the mass of dark matter particle for the spin-dependent
and spin-independent models. Also shown are the results from
the CMS Collaboration using the monophoton plus \MET channel~\cite{CMSMonophotons},
$\Pp\Pap$ collider experiment CDF~\cite{Aaltonen:2012jb},
direct detection experiments,  COUPP~\cite{bib:COUPP}, CoGeNT~\cite{bib:COGENT}, Picasso~\cite{bib:PICASSO}, XENON100~\cite{bib:XENON100},
CDMS II~\cite{bib:CDMSII2011,bib:CDMSII2010},  and SIMPLE~\cite{SIMPLE1},
and indirect  detection experiments,
IceCube~\cite{IceCube:2011aj} and Super-K~\cite{SUPERK}.
Table~\ref{tab:DM_limits} shows the 90\% CL limits on $\Lambda$ and the dark matter-nucleon cross section for the spin-dependent and spin-independent interactions.

\begin{table}
        \begin{center}
        \topcaption{
Observed 90\% CL limits on the dark matter-nucleon cross section and effective contact interaction scale $\Lambda$ for the spin-dependent and spin-independent interactions.
\label{tab:DM_limits}}
\begin{tabular}{c|cc|cc}
\hline
&\multicolumn{2}{c|}{ Spin-dependent }&\multicolumn{2}{c}{ Spin-independent}  \\
$M_{\chi}$ (\!\GeVcc)
&  $\Lambda$ (\GeVns)
& $\sigma_{\DM N}$ (cm$^{2}$)
&  $\Lambda$ (\GeVns)
& $\sigma_{\DM N}$ (cm$^{2}$)   \\
\hline
0.1         &  754  &   $ 1.03\times 10^{-42} $  &   749       &  $   2.90\times 10^{-41}$         \\
1            &  755  &   $ 2.94\times 10^{-41}$   &   751       &  $   8.21\times 10^{-40}$         \\
10          &  765  &   $ 8.79\times 10^{-41} $  &   760      &  $   2.47\times 10^{-39}$         \\
100        &  736  &   $ 1.21\times 10^{-40} $  &   764      &  $   2.83\times 10^{-39}$         \\
200        &  677  &   $ 1.70\times 10^{-40} $  &   736      &  $   3.31\times 10^{-39}$         \\
300        &  602  &   $ 2.73\times 10^{-40} $  &   690      &  $   4.30\times 10^{-39}$         \\
400        &  524  &   $ 4.74\times 10^{-40} $  &   631      &  $   6.15\times 10^{-39}$         \\
700        &  341  &   $ 2.65\times 10^{-39} $  &   455      &  $   2.28\times 10^{-38}$         \\
1000      &  206  &   $ 1.98\times 10^{-38} $  &   302      &  $   1.18\times 10^{-37}$         \\
\hline
\end{tabular}
\end{center}
\end{table}

\begin{table}
        \begin{center}
        \topcaption{
Observed and expected 95\% CL lower limits on the ADD model parameter $\MD$
(in \TeVcc) as a function of $\delta$, with and without NLO
$K$-factors applied.
\label{tab:STAT_ADD_limits}}

\begin{tabular}{c|cc|cc}
\hline
&\multicolumn{2}{c|}{ LO }&\multicolumn{2}{c}{ NLO }  \\
$\delta$& Exp. Limit  &  Obs. Limit  & Exp. Limit &  Obs. Limit \\
             & (\!\TeVcc)         &   (\!\TeVcc) & (\!\TeVcc)   &   (\!\TeVcc)\\ \hline
2   &           3.81    &  4.08       &       4.20 &        4.54 \\
3   &           3.06    &  3.24       &       3.32 &        3.51 \\
4   &           2.69    &  2.81       &       2.84 &        2.98 \\
5   &           2.44    &  2.52       &       2.59 &        2.71 \\
6   &           2.28    &  2.38       &       2.40 &        2.51 \\
\hline
\end{tabular}
\end{center}
\end{table}

\begin{figure}
  \begin{center}
  \includegraphics[scale=0.42] {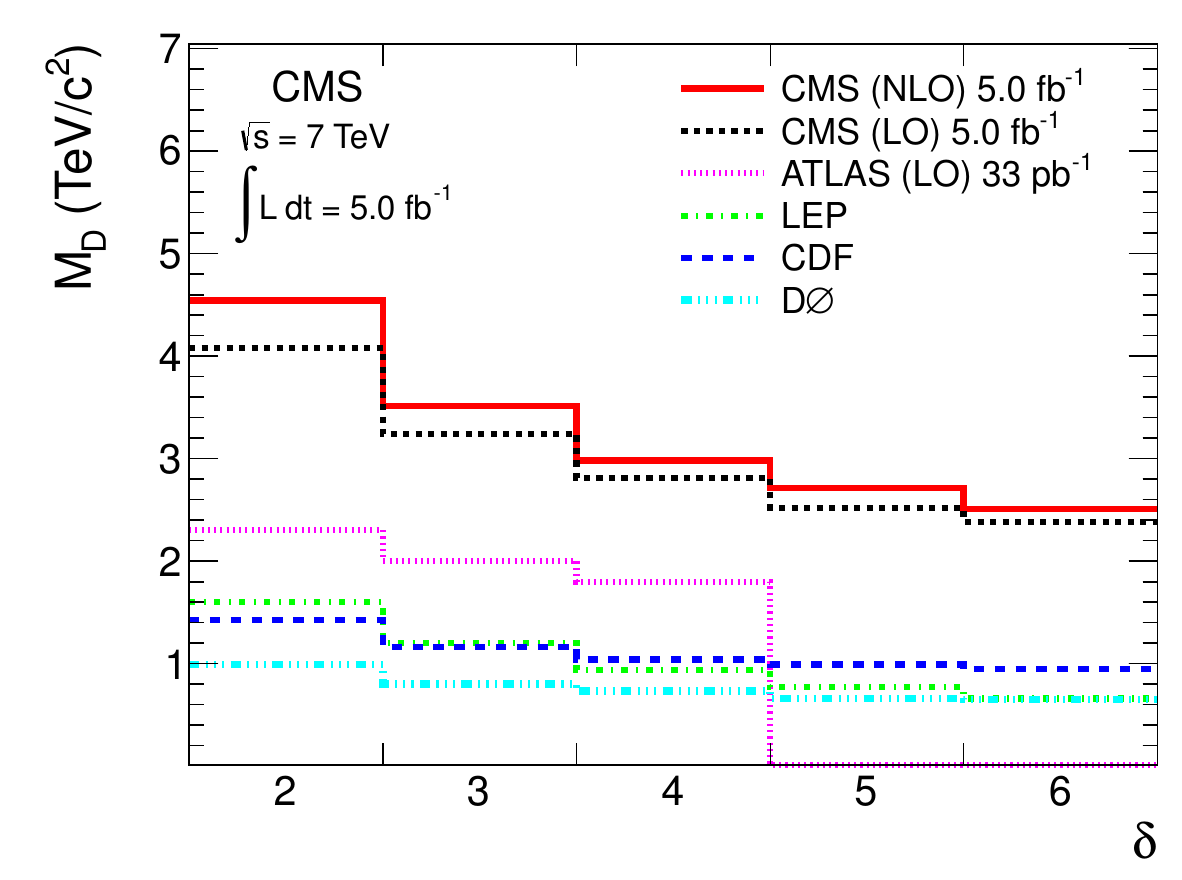}
   \caption{ Comparison of lower limits on $M_D$ versus the number of extra dimensions with ATLAS~\cite{bib:ATLASMonoJet}, LEP~\cite{bib:ALEPH,bib:OPAL,bib:DELPHI,bib:L3},
CDF~\cite{bib:CDFMonoPhoton}, and D0~\cite{bib:D0MonoPhoton}. 
\label{fig:ADD_limits_MD}}
  \end{center}
\end{figure}

Exclusion limits at 95\% CL for the large extra dimension ADD model parameter $\MD$ as a function of the number of extra dimensions are given in Table~\ref{tab:STAT_ADD_limits}.
A comparison of these results with results from previous searches is
shown in Fig.~\ref{fig:ADD_limits_MD}.  These limits are an
improvement over the previous best limits, by ${\sim}2\TeVcc$ for $\delta=2$ and
0.7\TeVcc for $\delta=6$.

\section{Summary}
A search has been performed for signatures of new physics yielding an
excess of events in the monojet and \MET channel. The results have
been used to constrain the pair production of dark matter particles in
models with a heavy mediator, and large extra dimensions in the
context of the
Arkani-Hamed, Dimopoulos, and Dvali
model. The data sample corresponds to an integrated luminosity of 5.0\fbinv and includes events containing a jet with transverse momentum above 110\GeVc
and \MET above 350\GeVc. Many standard model processes also have the same signature.
The QCD multijet contribution is reduced by several orders of magnitude to a negligible level using topological selections. The dominant backgrounds,
\ZnunuJets and $\PW$+jets, are estimated from data samples enriched in \Zmumu and \Wmunu events.
The data are found to be in good agreement with the expected contributions
from standard model processes.

A dark matter-nucleon scattering cross section in the framework of an effective theory is excluded above
$1.03\times 10^{-42}\ (1.21\times 10^{-40})\unit{cm}^2$ and
$2.90\times 10^{-41}\ (2.83\times 10^{-39})\unit{cm}^2$  for a dark matter particle
with mass 0.1 (100)\GeVcc at the 90\% CL  for the spin-dependent and spin-independent models, respectively.
For the spin-independent model, these are the best limits for dark
matter particles with mass  below 3.5\GeVcc, a region as yet
unexplored by the direct detection experiments.  For the
spin-dependent model, these limits represent the  most stringent constraints over the 0.1--200\GeVcc  mass range.

Values for the large extra dimensions
ADD
model parameter $\MD$ smaller than
4.54, 3.51, 2.98, 2.71, and 2.51\TeVcc
are excluded for a number of extra dimensions $\delta = $2, 3, 4, 5, and 6, respectively,
representing a significant improvement (1\TeVcc) over the previous limits.

\section*{Acknowledgments}

\hyphenation{Bundes-ministerium Forschungs-gemeinschaft Forschungs-zentren}We thank R. Harnik, P. J. Fox, and J. Kopp for the help in modeling dark matter production.
 We congratulate our colleagues in the CERN accelerator departments for the excellent performance of the LHC machine. We thank the technical and administrative staff at CERN and other CMS institutes, and acknowledge support from: FMSR (Austria); FNRS and FWO (Belgium); CNPq, CAPES, FAPERJ, and FAPESP (Brazil); MES (Bulgaria); CERN; CAS, MoST, and NSFC (China); COLCIENCIAS (Colombia); MSES (Croatia); RPF (Cyprus); MoER, SF0690030s09 and ERDF (Estonia); Academy of Finland, MEC, and HIP (Finland); CEA and CNRS/IN2P3 (France); BMBF, DFG, and HGF (Germany); GSRT (Greece); OTKA and NKTH (Hungary); DAE and DST (India); IPM (Iran); SFI (Ireland); INFN (Italy); NRF and WCU (Korea); LAS (Lithuania); CINVESTAV, CONACYT, SEP, and UASLP-FAI (Mexico); MSI (New Zealand); PAEC (Pakistan); MSHE and NSC (Poland); FCT (Portugal); JINR (Armenia, Belarus, Georgia, Ukraine, Uzbekistan); MON, RosAtom, RAS and RFBR (Russia); MSTD (Serbia); SEIDI and CPAN (Spain); Swiss Funding Agencies (Switzerland); NSC (Taipei); TUBITAK and TAEK (Turkey); STFC (United Kingdom); DOE and NSF (USA).
Individuals have received support from the Marie-Curie programme and the European Research Council (European Union); the Leventis Foundation; the A. P. Sloan Foundation; the Alexander von Humboldt Foundation; the Belgian Federal Science Policy Office; the Fonds pour la Formation \`a la Recherche dans l'Industrie et dans l'Agriculture (FRIA-Belgium); the Agentschap voor Innovatie door Wetenschap en Technologie (IWT-Belgium); the Council of Science and Industrial Research, India; the Compagnia di San Paolo (Torino); and the HOMING PLUS programme of Foundation for Polish Science, cofinanced from European Union, Regional Development Fund. 

\bibliography{auto_generated}
\cleardoublepage \appendix\section{The CMS Collaboration \label{app:collab}}\begin{sloppypar}\hyphenpenalty=5000\widowpenalty=500\clubpenalty=5000\textbf{Yerevan Physics Institute,  Yerevan,  Armenia}\\*[0pt]
S.~Chatrchyan, V.~Khachatryan, A.M.~Sirunyan, A.~Tumasyan
\vskip\cmsinstskip
\textbf{Institut f\"{u}r Hochenergiephysik der OeAW,  Wien,  Austria}\\*[0pt]
W.~Adam, E.~Aguilo, T.~Bergauer, M.~Dragicevic, J.~Er\"{o}, C.~Fabjan\cmsAuthorMark{1}, M.~Friedl, R.~Fr\"{u}hwirth\cmsAuthorMark{1}, V.M.~Ghete, J.~Hammer, N.~H\"{o}rmann, J.~Hrubec, M.~Jeitler\cmsAuthorMark{1}, W.~Kiesenhofer, V.~Kn\"{u}nz, M.~Krammer\cmsAuthorMark{1}, I.~Kr\"{a}tschmer, D.~Liko, I.~Mikulec, M.~Pernicka$^{\textrm{\dag}}$, B.~Rahbaran, C.~Rohringer, H.~Rohringer, R.~Sch\"{o}fbeck, J.~Strauss, A.~Taurok, W.~Waltenberger, G.~Walzel, E.~Widl, C.-E.~Wulz\cmsAuthorMark{1}
% ...
\vskip\cmsinstskip
\dag:~Deceased\\
1:~~Also at Vienna University of Technology, Vienna, Austria\\
% ...

\end{sloppypar}
\end{document}